# Deep Learning-based Four-region Lung Segmentation in Chest Radiography for COVID-19 Diagnosis

Young-Gon Kim[1], Kyungsang Kim[2], Dufan Wu[2], Hui Ren[2], Won Young Tak[3], Soo Young Park[3], Yu Rim Lee[3], Min Kyu Kang[4], Jung Gil Park[4], Byung Seok Kim[5], Woo Jin Chung[6], Mannudeep K. Kalra[2], Quanzheng Li[2]*

**Authors' institutions**

[1] *Department of Pathology, Seoul National University Hospital, Seoul, South Korea*

[2] *Department of Radiology, Massachusetts General Hospital, Boston, MA, USA*

[3] *Department of Internal Medicine, School of Medicine, Kyungpook National University, Daegu, South Korea*

[4] *Department of Internal Medicine, Yeungnam University College of Medicine, Daegu, South Korea*

[5] *Department of Internal Medicine, Catholic University of Daegu School of Medicine, Daegu, South Korea*

[6] *Department of Internal Medicine, Keimyung University School of Medicine, Daegu, South Korea*

*\*E-mail:* li.quanzheng@mgh.harvard.edu

## Abstract

**Purpose.** Imaging plays an important role in assessing severity of COVID-19 pneumonia. However, semantic interpretation of chest radiography (CXR) findings does not include quantitative description of radiographic opacities. Most current AI-assisted CXR image analysis framework do not quantify for regional variations of disease. To address these, we proposed a four-region lung segmentation method to assist accurate quantification of COVID-19 pneumonia.

**Methods.** A segmentation model to separate left and right lung is firstly applied, and then a carina and left hilum detection network is used, which are the clinical landmarks to separate the upper and lower lungs. To improve the segmentation performance of COVID-19 images, ensemble strategy incorporating five models is exploited. Using each region, we evaluated the clinical relevance of the proposed method with the Radiographic Assessment of the Quality of Lung Edema (RALE).

**Results.** The proposed ensemble strategy showed dice score of 0.900, which is significantly higher than conventional methods (0.854-0.889). Mean intensities of segmented four regions indicate positive correlation to the extent and density scores of pulmonary opacities under the RALE framework.

**Conclusion.** A deep learning-based model in CXR can accurately segment and quantify regional distribution of pulmonary opacities in patients with COVID-19 pneumonia.

## Abbreviations

CXR = chest X-ray radiography, RALE = Radiographic Assessment of the Quality of Lung Edema, ARDS = Acute Respiratory Distress Syndrome, RT-PCR = Reverse Transcription Polymerase Chain Reaction, RUR = Right Upper Region, RLR = Right Lower Region, LUR = Low Upper Region, LLR = Left Lower Region, mAP = mean of Average Precision

**Summary**

The proposed algorithm that consists of two deep learning-based models for detection of left hilum and segmentation of left and right lung regions can divide a whole lung into four-regions, i.e., LUR, RUR, LLR and RLR, in chest radiography for COVID-19 patients.

**Key Points**

- A proposed algorithm divided the whole lung region into four-regions, LUR, RUR, LLR and RLR, and mean intensity within each region showed positive correlation with extent and density scores of pulmonary opacities from radiologists.
- A majority voting-based ensemble method and augmentation methods enhanced segmentation model performance.
- The proposed algorithm can have potential to be widely adopted as the first step for analysis of lung regions in chest radiography for COVID-19 patients.

**Introduction**

The COVID-19 is a novel infectious disease, caused by severe acute respiratory syndrome coronavirus 2 (SARS-CoV-2), which could lead to acute respiratory distress syndrome (ARDS) [1, 2]. Starting in December 2019 from the province of Hubei, China, COVID-19 became a pandemic that has claimed over 800,000 lives, infected over 24 million people worldwide, and wrecked economic and social hardships in all six inhabited continents [3]. Real-time reverse transcription polymerase chain reaction (RT-PCR) is the preferred test for confirming COVID-19 infection. Despite its limitations and limited availability in several parts of both developed and developing world, most international and national organizations recommend RT-PCR assays for screening and initial diagnosis of COVID-19 infection.

Use of imaging, computed tomography (CT) and chest radiography (CXR), for initial diagnosis of COVID-19 pneumonia is extremely common in sites with high prevalence and/or limited availability of RT-PCR assays. However, there is consensus that imaging should be used judiciously, and most often, in patients with moderate to severe disease and those with complications and comorbidities. Both CT and CXR are used for establishing disease extent or severity of pulmonary opacities. Compared to CT, CXRs are more accessible, mobile, cheaper, lower dose, efficient, as well as easier to sanitize and use in intensive care settings. Prior studies have reported on role of these imaging modalities for initial diagnosis and qualitative severity of COVID-19 pneumonia [2, 4, 5].

For automated classification and detection of COVID-19 cases, deep learning-based methods with both CT and CXRs have been proposed [6-9]. To assess disease severity from quantitative extent of pneumonia, an automatic method for prediction of severity score have been introduced with a deep learning-based method [10], which showed high correlation score at $R^2$ 0.865 and 0.746 with radiological extent and opacity. Some clinical studies using the CXR have tried to segment the whole lung into subsets for predicting the severity of the diseases [11]. With importance of early diagnose in CXRs, lung segmentation methods have been used to reduce non-specific signals such as tube or lines efficiently [12, 13].

CXR findings in patients with COVID-19 pneumonia range from normal lungs and subtle haziness in mild or early to more extensive diffuse opacities consistent with diffuse pnueumonia and adult respiratory distress syndrome (ARDS) in severe and advanced disease. Radiographic assessment of lung edema (RALE) is a score indicating the severity of ARDS and COVID-19 pneumonia on CXRs [14]. For RALE score, each lung is divided into two regions, upper and lower based on a horizontal line through the level of origin of the left upper lobar bronchus from the left mainstem bronchus. Then, density and extent of pulmonary opacities is subjectively graded by radiologists in each of the four regions to determine the regional and total scores of pulmonary opacities. The RALE score has been validated as a good predictor for ARDS [11]. However, the method is prone to inter- and intra-observer variations, challenging in settings of low lung volume, and too tedious and inefficient for incorporation into interpretation routine. Other studies have proposed six-region division of lungs [15, 16].

In this study, we propose a deep learning-based model to segment four regions of lung in CXRs of COVID-19 patients. To achieve a robust four-region lung segmentation, two deep learning-based segmentation and detection models are proposed as shown in Fig. 1. For the four-region lung mask, left and right regions are firstly segmented, where a majority voting ensemble method is used from five deep learning-based segmentation models. Then, the upper and lower sub regions are divided by the positions of carina and hilum predicted by a deep learning-based detection model. For validation of the segmented regions, each mean intensity calculated by normalized pixels for each region is used to validate correlation with extent and density scores of pulmonary opacities.

**Materials and methods**

*Data Description*

　*1) Segmentation.* Since anatomic segmentation of lungs is independent of radiographic abnormalities, for training segmentation models, two public datasets; RSNA pneumonia detection challenge dataset [17] and JSRT dataset [18], were used. RSNA pneumonia detection challenge dataset consists of 568 CXRs from tuberculosis chest dataset in department of health and human services (HHS) of Montgomery county and JSRT dataset consists of 257 CXRs from JSRT dataset in Japanese society of radiological technology (JSRT) in cooperation with the Japanese Radiological Society, were used to train segmentation models.

　For evaluation of the segmentation model performance, 200 CXRs of 51 patients with COVID-19 pneumonia were obtained from three hospitals in South Korea including Kyungpook National University Hospital, Daegu Catholic University Hospital, and Yeungnam University Hospital.

　*2) Detection.* The carina and left hilum detection algorithms were trained on another 704 CXRs from 166 patients with confirmed COVID-19 pneumonia between February-May 2020, at the same hospitals in South Korea, including Kyungpook National University Hospital, Daegu Catholic University Hospital, and Yeungnam University Hospital (Table I). The positions of carina and left hilum were annotated under the supervision of a subspecialty chest radiologist with 13 years of clinical experience in thoracic imaging. For each CXR, a bounding box was placed around the left hilum. The inferior margin of carina was also annotated with a point marker. A bounding box of 100 pixels centered at the carina point was used for the training of carina detection algorithm.

　*3) Correlation.* To further validate the proposed 4-region segmentation algorithm, each CXR was evaluated for its RALE score. The RALE score was evaluated by giving extent (0-4) and density (0-3) scores of pulmonary opacities in each region of the lung [11]. For each region, the correlation between its mean intensity and the corresponding extent and density scores of pulmonary opacities were analyzed.

*Segmentation model*

U-net architecture [19] using skip connection was selected to train the segmentation models, which is the most widely used network structure for segmentation in the field of medical imaging. We trained five segmentation models with different conditions including backbones, pre-processing, and augmentation properties as shown in Table 1. EfficientNet v0 and v7 architecture [20] were used as the backbone network in the U-net to train the first to third segmentation models and the fourth and fifth segmentation models, respectively. Gaussian noise and gamma correction were adjusted to improve the robustness of the models to pixel noises from the portable devices. To train segmentation models robust to Anterior-Posterior (AP) CXRs that is not included in the public datasets, morphological transformation methods such as grid distortion, affine transform, and elastic transformation with different parameters were used as augmentation methods [21]. Five binary masks were used to generate an ensemble mask based on the majority voting method. Technically, if a half of masks were predicted as a lung for a pixel, the pixel is labeled as a lung region.

In addition, post-processing steps were taken to refine the ensemble mask. All the holes were filled with the dilation operation and the isolated regions were eliminated.

The augmentation methods with different parameters were adjusted during training [5] in the five models. All models were trained with same hyper-parameters, such as Adam optimizer (learning rate: 0.0001), epochs (200), batch size (8) and same input size at $256 \times 256$. Best models were selected at the lowest loss on the validation dataset.

*Detection model*

We propose a novel and robust method to find a central point for segmentation of the whole region into four-regions such as right upper region (RUR), right lower region (RLR), low upper region (LUR), and left lower region (LLR). Although conventional RALE score described a horizontal line through the origin of the left upper lobe bronchus for 4-segment classification of lungs, it is difficult to see this point in most patients with COVID-19 with portable CXRs. As a surrogate, the left hilum is the closest landmark for dividing upper and lower regions. However, the left hilum is sometimes difficult to be detected in those patients with advanced disease or patient rotation. On the other hand, carina is clear under most circumstances, and its relative position to the left hilum is stable at approximately 2cm [22] above the left hilum vertically. Therefore, we also used carina to identify the central point for horizontal lung segmentation into upper and lower regions.

RetinaNet [23] was used to train the detection model for the carina and the left hilum. The central point of prediction box is used as a reference horizontal level that divide the upper and lower lungs. Most of the time, we select the prediction box for the left hilum for dividing the lung into upper and lower regions. However, if the model confidence of the left hilum detection is lower or equal to 0.9, the prediction box for the carina would be used.

To train a robust detection model, augmentation methods [21] such as rotation, translation, shearing, scaling, pixel noise, different range of contrast, brightness, hue, and saturation were used. The best model was selected as the lowest total loss in the validation set.

*Normalization*

Intensity normalization is normally used as a pre-processing to reduce statistical distribution of the intensity among input CXRs. Different devices or setting parameters cause CXRs showing brightness differences as shown in Fig. 2. Density scores of Fig. 2(a)-(d) were confirmed at 0 while each showed quite different mean intensities of the lung at 39.8, 34.4, 16.6, and 13.2, respectively. To reduce this variation, intensity normalization was conducted. Pixels inside of the lung were normalized by subtracting their values with the mean intensities outside of the lung regions. The normalized pixels were averaged to obtain each representative value for each region to evaluate its correlation with extent and density scores of pulmonary opacities.

*Correlation with RALE score*

Extent (0-4) and density (0-3) scores of pulmonary opacities were manually assigned by an experienced radiologist for each region of the lung according to the guideline [11].

The extent and density scores of pulmonary opacities were correlated with the mean intensity corresponding to the same location divided by the proposed algorithm. To evaluate if there is a linear relationship between regional mean intensity and the RALE score, we used the subset of COVID-19 patients with a RALE score larger than 0. Pearson correlation [24] was used to test the relationship.

*Statistical Evaluation*

Model performance comparisons for segmentation were conducted with anonymized dataset (three hospitals in South Korea) in terms of Dice score to select the best segmentation model. Then, we conducted pair-wise comparisons of Dice scores between the ensemble model and others to show significant difference ($p < 0.05$).

*Experimental Environment*

Experimental environments were on Ubuntu 16.04 with a Tesla V-100 GPU, CUDA 9.0/cuDNN 7.0 (NVidia Corporation), and Keras 2.0 deep learning platform.

**Results**

Model performance for segmentation are listed in Table 3. The first to the fifth segmentation models were merged to the ensemble model. Model performance of the ensemble model including all models had the highest dice coefficient (0.908 ± 0.057) with significant statistical differences from Model 1 to 5 (All $p<0.05$).

Fig. 3 shows an example of advantages of the ensemble method for different quality of CXRs. The first to the last row in each column shows an input CXR, the ground truth mask, the ensemble result, and the five results predicted by the individual segmentation models. Fig. 3(a-1) shows a high quality CXR without medical device, substantial patient rotation, and over- or under- radiographic exposure. The five individual models gave consistent segmentations shown in Fig. 3(a-4)-(a-8).

The CXR in Fig. 3(b-1) was challenging due to consolidation and/or atelectasis in the left lower lobe which obscures delineation of left lung hilum. Left lung hilum can also be obscured by dense perihilar opacities or marked patient rotation. Compared to the consistent results predicted by the first to third models as shown in Fig. 3(b-4)-(b-6) (0.929, 0.943, 0.934), the masks resulting from model 4 and 5 under-estimated the area of right lung (0.817, 0.831). The ensemble could still reach a robust mask (0.933) as shown in Fig. 3(b-3).

Fig. 3(c-1) shows a left chest tube traveling up to and obscuring visualization and detection of left hilum. Compared to the consistent results predicted by the third to fifth models as shown in Fig. 3(c-6)-(c-8) (0.883, 0.879, 0.903), the first and second models labeled areas outside of lung regions as shown in Fig. 3(c-4) and Fig. 3(c-5) (0.783, 0.885) due to extending into the right chest wall subcutaneous emphysema which has intensity similar to the right lung. The ensemble results gave a relative robust mask (0.899) as shown in Fig. 3(c-3).

Model performance for detection of left hilum and carina in terms of mean of average precision (mAP [37]) was observed at 0.694. Fig. 4 shows different examples for selection of a reference point to divide upper and lower lungs. Fig. 4(a) shows an example with high confidence in detection result (left hilum: 0.94), where the center of the left hilum bound box is directly used as the reference horizontal level for the upper and lower lung region separation as shown in Fig. 4(b). In Fig. 4(c), the confidence of the detection result was low (left hilum: 0.56), and then the vertically 2cm lower position of the carina bound box was used for the upper and lower region separation as shown in Fig. 4(d).

After normalization, the mean intensity of each region was correlated with the corresponding extent (0-4) and opacity (0-3) scores. Fig. 5(a)-(d) shows the correlation of the extent score with mean intensities for each region, i.e., RUR, LUR, RLR, and LLR. For each region, the mean intensity increased as the extent scores increased. The correlation with the extent score for the LUR showed a strong positive linear relationship at 0.716 (<0.001) as shown in Fig. 5(c), and correlation values with the extent score for LUR, RUR, and RLR showed moderate positive linear relationship at 0.625 (<0.001), 0.454 (<0.001), and 0.457 (<0.001), respectively, as shown in Fig. 5(a), (b), and (d).

In case of density scores, the tendency that each mean intensity increased as the density scores increased was observed as shown in Fig. 5(e)-(h). The correlation with the density scores for RUR, LUR, RLR, and LLR showed moderate positive linear relationship at 0.553 (<0.001), 0.469 (<0.001), 0.506 (<0.001), and 0.465 (<0.001), respectively.

Distribution of mean intensity for each region is shown in Fig. 6. Sum of left lung region is higher than that of right lung region. The mean intensity of LLR where heart is not segmented in the segmentation algorithm is lower than that of other regions.

**Discussion**

In this work, we proposed a four-lung region auto-segmentation algorithm and validated the algorithm with correlation of the mean intensities for each region segmented by the algorithm with extent and density scores of pulmonary opacities from the radiologist.

In the detection model for the carina and left hilum, the mAP for the carina was higher at 0.743 than that for the left hilum at 0.467 since the quality of labeling the left hilum was inferior to that of the carina. It is because labeling the exact locations of the left hilum is harder due to wider longitudinal extent, overlap from cardiomediastinal structures, obscuration from adjacent pulmonary opacities, or overlapping lines and tubes, while labeling the location of the carina is easier due to less noise. It showed that the model performance highly depends on the quality of labeling data though same number of training set with labeling data were used.

In validation of the segmentation models, two different public datasets (RSNA pneumonia detection challenge and JSRT datasets) and the anonymized dataset were used. On the public datasets, the five individual models showed comparable dice coefficients at around 0.958-0.967 due to high radiographic quality of the posterior-anterior (PA) projection public datasets with low disease burden; both factors make it easier for each model to segment lung regions. However, in the testing dataset most CXRs for COVID-19 patients were captured with AP CXR instead of the PA CXRs in the training dataset. The AP CXRs were limited due to lower radiographic quality, lower lung volumes, patient rotation, and a larger number of chest tubes, lines, and devices. To overcome these issues, the ensemble method was selected with segmentation models trained with different conditions. In training segmentation model, different augmentation properties and backbone networks with ensemble lead to robust lung mask in different situation such as position, image quality, intubated patients, etc. To train the segmentation model robust to noise from portable device and posture that never been seen when training, Gaussian noise and distortion transform augmentations were used. Different backbone networks were adjusted due to train models robust to intubated patients and low contrast CXRs, which predicted robust lung masks.

Correlation of the extent and density scores of pulmonary opacities with the mean intensities for each lung region showed at least moderate positive linear relationship as shown in Fig. 3. For RLR, the correlation of extent score with a mean intensity showed a strong positive relationship at Pearson correlation 0.716 ($p<0.005$). Apart from the inherent limitations of portable AP projection CXRs, the less than substantial correlation between the model and the subjective scores can also be related to type of pulmonary opacities. Dense basilar opacities in COVID-19 pneumonia, likely related to severe airspace opacification (or consolidation on CT images), obscure the lung margins at their interface with hemidiaphragm and cardiomediastinal structures (note the obscured lower lungs in Fig 4(a)) and consequent underestimation of lung volume with segmentation as well as evaluation of the extent and density of pulmonary opacities. Such opacities require dedicated training datasets which were not available to our model.

Although we showed the Pearson correlation of the segmented regions with extent and density scores of pulmonary opacities, the proposed method still has a great potential combined with various clinical applications. For the development of clinical methods, the segmentation model will be a crucial pre-processing tool for extracting lung regions in CXRs. In the future, combined with the proposed method as a pre-clinical step, we will develop an automatic prediction method of the RALE and the severity prediction model for COVID-19 patients.

**Conclusion**

In this paper, we proposed the deep learning-based four-region lung segmentation method in CXRs for COVID-19 patients, where the detection model to find the center positions of carina and hilum structures was incorporated to divide upper and lower regions. The proposed ensemble method based on five segmentation models trained with different augmentations and backbone networks showed significantly high performance than a single model in terms of dice coefficient. To evaluate the feasibility of the proposed method, we confirmed the positive correlation between intensities of segmented regions and extent and density scores of pulmonary opacities. Future work will focus on automatic prediction of the RALE and clinical evaluations using CXRs from multiple sites and the severity prediction model for COVID-19 patients.

**Author Contributions**

**Y.-G.K.**, **K.K., D.W.,** and **H.R.** analyzed data, searched literature, generated figures, and interpreted data. **Q.L.** designed and supervised the study. **W.Y.Y., S.Y.P., Y.R.L., M.K.K., J.G.P., B.S.K.**, **W.J.C.**, and **M.K.K.** labeled data. All the authors were involved in writing the paper and had final approval of the submitted and published versions.

**Disclosures of Conflicts of Interest:**

The authors declare no competing interests.

**Code availability**

https://github.com/younggon2/Research-Segmentation-Lung-CXR-COVID19

**11**(2): p. 125.
22. Chassagnon, G., et al., *Tracheobronchial branching abnormalities: lobe-based classification scheme.* Radiographics, 2016. **36**(2): p. 358-373.
23. Lin, T.-Y., et al. *Focal loss for dense object detection.* in *Proceedings of the IEEE international conference on computer vision.* 2017.
24. Benesty, J., et al., *Pearson correlation coefficient*, in *Noise reduction in speech processing*. 2009, Springer. p. 1-4.
17

**Figure Legends**

Fig. 1. A flowchart of the proposed algorithm for segmentation of zones of the lung in CXR of COVID-19 patient. Right (R) and left (L) lung masks are generated by an ensemble method based on the majority voting from five lung masks predicted by models trained with different conditions. Then, left hilum and carina are detected and used to find a central point to split the whole lung into upper and lower regions. Finally, right upper lung (RUR), right lower lung (RLR), low upper lung (LUR), and left lower lung (LLR) are obtained.

Fig. 2. Cases with same density scores (0) but with distinct mean intensities on CXR. Mean intensities of lungs from (a) to (d) are 39.8, 34.4, 16.6, and 13.2, respectively.

Fig. 3. An example of advantages of the ensemble method for different quality of CXRs. The first to last row in each column shows an input CXR, a ground truth mask, an ensemble result, and five results predicted by the first to fifth model. (a-1) A clear CXR that shows none of severe noise from a portable device and obstacles like medical devices, (b) a lung mask of (a-1), (a-3) an ensemble mask from the first to the fifth masks (a-4)-(a-8). Dice coefficients of (a-3)-(a-8) are 0.955, 0.928, 0912, 0.948, 0.948, and 0.948, respectively. (i) An CXR showing severe blurry within both lung regions due to lung opacity, (b-2) a lung mask of (b-1), (b-3) an ensemble mask from the first to the fifth masks (b-4)-(b-8). Dice coefficients of (b-3)-(b-8) are 0.955, 0.928, 0912, 0.948, 0.948, and 0.948, respectively. (c-1) An CXR showing sever noise generated from a portable device, (c-2) a lung mask of (c-1), (c-3) an ensemble mask from the first to the fifth masks (c-4)-(c-8). Dice coefficients of (c-3)-(c-8) are 0.899, 0.783, 0.885, 0.883, 0.879, and 0.903, respectively.

Fig. 4. An example of detection results for the left hilum colored at red and carina colored at green and, dividing segmented lung mask into upper and lower lungs, i.e., RUR, LUR, RLR, and LLR with a reference point colored at while. (a) detection results for the left hilum (confidence: 0.94) and the carina (0.98). (b) A center point of the detection box for the left hilum is used as the reference point to divide upper and lower lungs. (c) detection results for the left hilum (0.56) and carina (0.95). (d) A location down to approximately 2cm vertically from a center point of the detection box for the carina is used as the reference point to divide upper and lower lungs.

Fig. 5. Boxplots of mean intensities with extent scores (0-4) and density scores (0-3) of pulmonary opacities for four-regions. (a) and (e) RUR, (b) and (f) LUR, (c) and (g) RLR, (d) and (h) LLR. For each region, the mean intensity increased as the extent and density scores increased.

Fig. 6. Boxplots of mean intensities for four-regions. The mean intensity of LLR where heart is not segmented in the segmentation algorithm is lower than that of other regions.

**Tables**

Table 1. Conditions for training different segmentation models.

| Model   | Backbone  | Pre-proc. | Augmentation |
|---------|-----------|-----------|--------------|
| Model 1 | Efficient0 | N/A      | DA |
| Model 2 | Efficient0 | HE       | DA |
| Model 3 | Efficient0 | HE       | DA + Gaussian noise (0.5) + gamma correction (0.5) + grid distortion (0.1) + elastic transform (0.1) + affine transform (0.1) |
| Model 4 | Efficient7 | HE       | DA + Gaussian noise (0.5) + gamma correction (0.5) |
| Model 5 | Efficient7 | HE       | DA + Gaussian noise (0.5) + gamma correction (0.5) + grid distortion (0.1) + elastic transform (0.1) + affine transform (0.1) |

Abbreviations: HN, histogram normalization; DA, default augmentation (horizontal flip: 0.5, rotation: 25°, random contrast: 0.1, random brightness 0.1, gamma correction: 0.1, Gaussian noise: 0.1, contrast limited adaptive histogram equalization 0.1.

Table 2. Demographics of the dataset for carina and left hilum detection.

|         | Training set (n = 551) | Validation set (n = 153) | Testing set (n = 162) |
|---------|------------------------|--------------------------|-----------------------|
| Patient | 124                    | 42                       | 42                    |
| Age     | 68.3 ± 14.8            | 59.5 ± 16.2              | 54.3 ± 18.4           |
| Male    | 53 (42.7%)             | 16 (38.0%)               | 23 (54.7%)            |
| RALE    | 9.9 ± 10.7             | 3.9 ± 6.7                | 4.2 ± 6.2             |
| Death   | 43 (34.6%)             | 2 (4.7%)                 | 4 (9.5%)              |

Table 3. Performance comparison with single and ensemble model in terms of dice coefficient for the anonymized dataset in South Korea.)

| No. | Model    | Mean ± Std.       |
|-----|----------|-------------------|
| 1   | Model 1  | 0.874 ± 0.057*    |
| 2   | Model 2  | 0.854 ± 0.072*    |
| 3   | Model 3  | 0.873 ± 0.089*    |
| 4   | Model 4  | 0.888 ± 0.084*    |
| 5   | Model 5  | 0.889 ± 0.079*    |
| 6   | Ensemble | 0.900 ± 0.074     |

(* Indicates a significant difference between an ensemble and other models, $p < 0.05$)

**Figures**

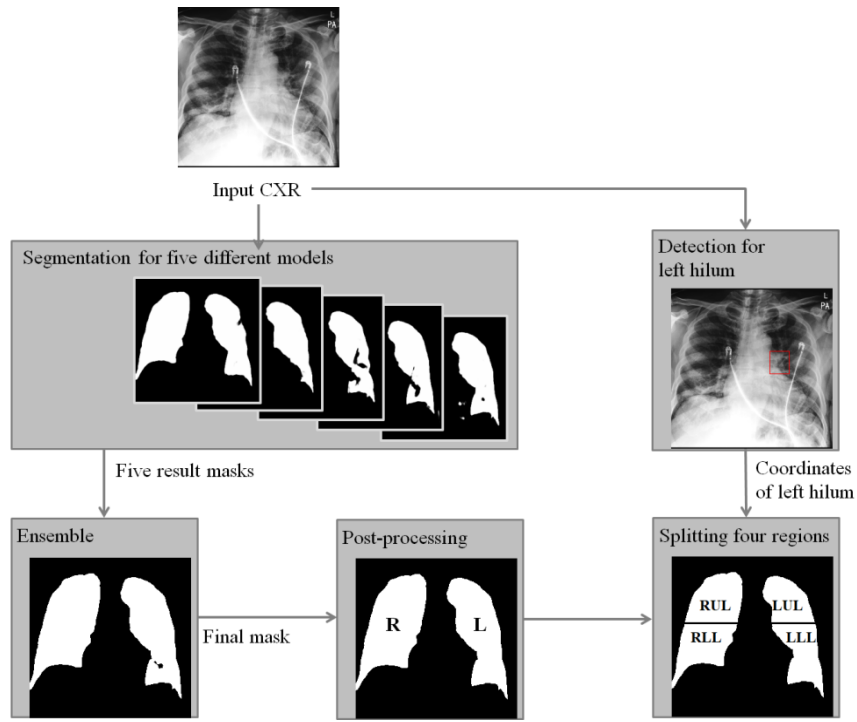

Fig. 1. A flowchart of the proposed algorithm for segmentation of zones of the lung in CXR of COVID-19 patient. Right (R) and left (L) lung masks are generated by an ensemble method based on the majority voting from five lung masks predicted by models trained with different conditions. Then, left hilum and carina are detected and used to find a central point to split the whole lung into upper and lower regions. Finally, right upper lung (RUR), right lower lung (RLR), low upper lung (LUR), and left lower lung (LLR) are obtained.

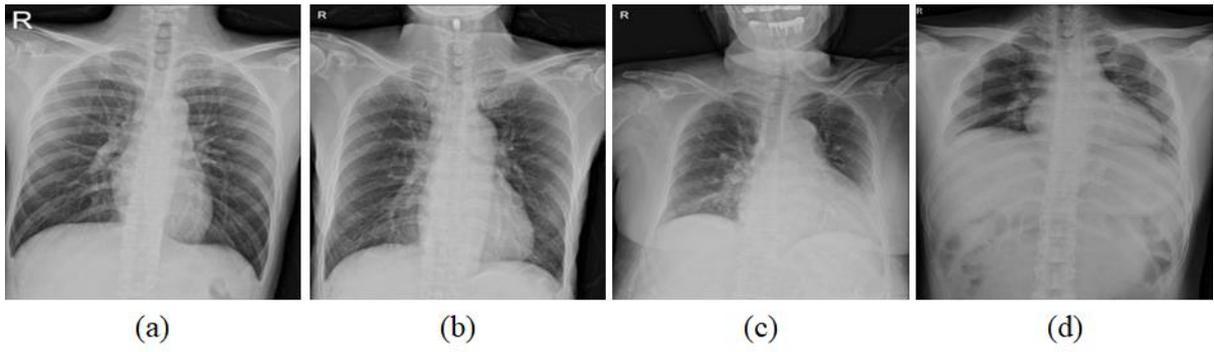

Fig. 2. Cases with same density scores (0) but with distinct mean intensities on CXR. Mean intensities of lungs from (a) to (d) are 39.8, 34.4, 16.6, and 13.2, respectively.

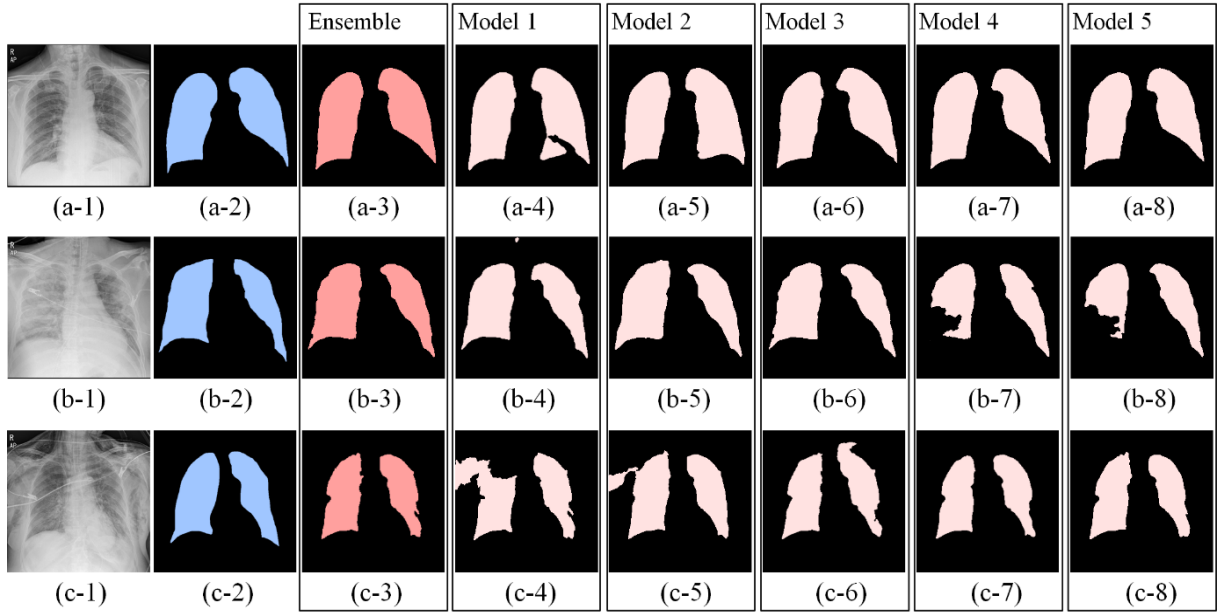

Fig. 3. An example of advantages of the ensemble method for different quality of CXRs. The first to last row in each column shows an input CXR, a ground truth mask, an ensemble result, and five results predicted by the first to fifth model. (a-1) A clear CXR that shows none of severe noise from a portable device and obstacles like medical devices, (b) a lung mask of (a-1), (a-3) an ensemble mask from the first to the fifth masks (a-4)-(a-8). Dice coefficients of (a-3)-(a-8) are 0.955, 0.928, 0912, 0.948, 0.948, and 0.948, respectively. (i) An CXR showing severe blurry within both lung regions due to lung opacity, (b-2) a lung mask of (b-1), (b-3) an ensemble mask from the first to the fifth masks (b-4)-(b-8). Dice coefficients of (b-3)-(b-8) are 0.955, 0.928, 0912, 0.948, 0.948, and 0.948, respectively. (c-1) An CXR showing sever noise generated from a portable device, (c-2) a lung mask of (c-1), (c-3) an ensemble mask from the first to the fifth masks (c-4)-(c-8). Dice coefficients of (c-3)-(c-8) are 0.899, 0.783, 0.885, 0.883, 0.879, and 0.903, respectively.

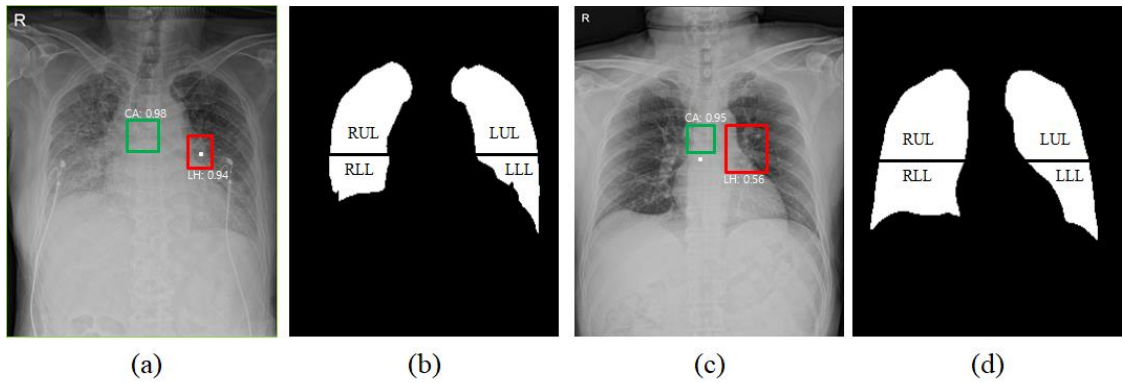

Fig. 4. An example of detection results for the left hilum colored at red and carina colored at green and, dividing segmented lung mask into upper and lower lungs, i.e., RUR, LUR, RLR, and LLR with a reference point colored at while. (a) detection results for the left hilum (confidence: 0.94) and the carina (0.98). (b) A center point of the detection box for the left hilum is used as the reference point to divide upper and lower lungs. (c) detection results for the left hilum (0.56) and carina (0.95). (d) A location down to approximately 2cm vertically from a center point of the detection box for the carina is used as the reference point to divide upper and lower lungs.

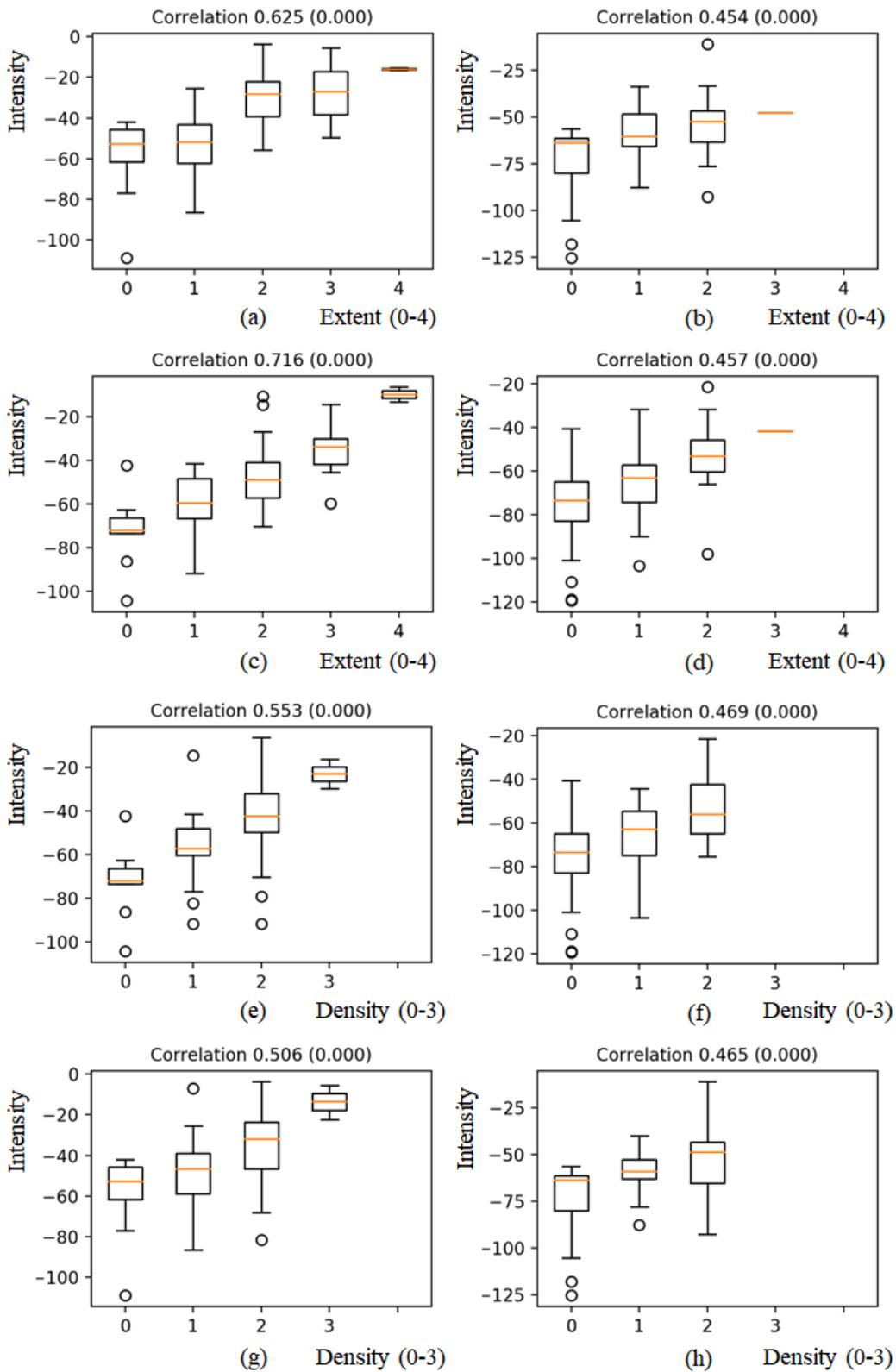

Fig. 5. Boxplots of mean intensities with extent scores (0-4) and density scores (0-3) of pulmonary opacities for four-regions. (a) and (e) RUR, (b) and (f) LUR, (c) and (g) RLR, (d) and (h) LLR. For each region, the mean intensity increased as the extent and density scores increased.

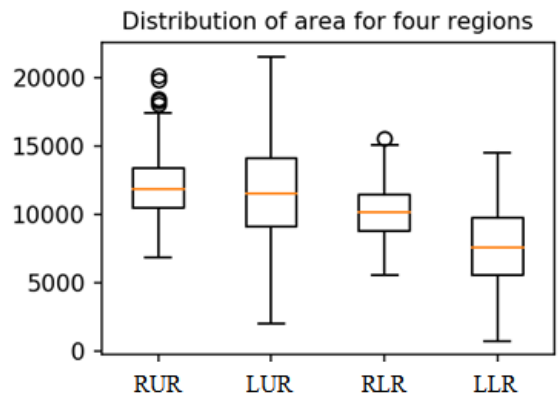

Fig. 6. Boxplots of mean intensities for four-regions. The mean intensity of LLR where heart is not segmented in the segmentation algorithm is lower than that of other regions.